# Growth modes of nanoparticle superlattice thin films

D. Mishra[1], D. Greving[1], G. A. Badini Confalonieri[1,2], J. Perlich[3], B.P. Toperverg[1,4], H. Zabel[1], and O. Petracic[1,5]

[1] Institute for Experimental Condensed Matter Physics, Ruhr-University Bochum, 44780 Bochum, Germany
[2] Instituto de Ciencia de Materiales, E-28049 CSIC Madrid, Spain
[3] Deutsches Elektronen-Synchrotron DESY, 22607 Hamburg, Germany
[4] Petersburg Nuclear Physics Institute, 188300 Gatchina, St. Petersburg, Russia
[5] Jülich Centre for Neutron Science JCNS and Peter Grünberg Institut PGI, JARA-FIT, Forschungszentrum Jülich GmbH, 52425 Jülich, Germany

**Abstract:** We report about the fabrication and characterization of iron oxide nanoparticle thin film superlattices. The formation into different film morphologies is controlled by tuning the particle plus solvent-to-substrate interaction. It turns out that the wetting vs. dewetting properties of the solvent *before* the self-assembly process during solvent evaporation plays a major role to determine the resulting film morphology. In addition to layerwise growth also three-dimensional mesocrystalline growth is evidenced. The understanding of the mechanisms ruling nanoparticle self-assembly represents an important step toward the fabrication of novel materials with tailored optical, magnetic or electrical transport properties.

The advent of controlled thin film growth about seven decades ago revolutionized many areas of science and technology such as optical coatings [1, 2], magnetic layers and multilayers [3, 4] or semiconductor thin films [5, 6]. In the early stage of research on thin films it soon became clear that it was imperative to understand the mechanisms which control and define the growth of thin films to gain control over the physical properties of these novel artificial materials. Hence huge efforts of the scientific community were dedicated to characterize, optimize and understand film growth. Thin films are evidently composed of atoms, which are considered as zero-dimensional building blocks. Extending this concept to the case of films composed of nanoparticles, the assumption is made that *nanoparticles* (also termed *'nanocrystals'*) can also serve as zero-dimensional building blocks. By self-assembly, these building blocks may form two-dimensional thin films or three-dimensional crystals (so-called 'nanoparticle superlattices') analogous to atomic films and crystal lattices [7 - 17]. However, in contrast to atomic films, the solid state properties of those elementary building blocks (dielectric, magnetic, etc.), as well as their interactions responsible for the type of ordering, can be modified for achieving desirable characteristics of the resulting nanocomposed supercrystals.

Indeed, modern chemical synthesis methods have enabled the fabrication of nanoparticles (NPs) from metallic, semiconducting and insulating materials in various shapes and sizes, and with narrow size distributions [12, 18 - 21]. Moreover, there are considerable efforts to design novel self-assembled structures with tunable magnetic, electrical and optical properties for the fabrication of nanodevices or materials. Therefore, it is essential to understand the underlying mechanism of self-assembly in 2 or 3 dimensions. It is thus necessary to address the forces and interactions on the nanoscale responsible for the different growth modes of nanoparticle superlattices.

There are numerous superstructures or morphologies observed upon self-assembly, starting from monolayers with cubic, hexagonal or face centered cubic geometry [11, 12, 13], networks [22], rings [23], chains [24] and 3-dimensional islands [25]. A conscious effort has also been made to explain these morphologies by a comprehensive theoretical framework. One approach is to consider various interaction forces relevant at the nanoscale [24], namely, van der Waals (vdW) force, electrostatic force, magnetic dipole force, steric repulsion, depletion force and capillary force. The simplest Lennard-Jones potential, which includes vdW force, can describe the atomic crystallization process involving two body interactions summed over all pairs participating in the crystal formation. However, thin films and superlattices are usually grown from NP colloids on some kind of substrate. Therefore, in self-assembly of nanoparticles the solvent plus NP (one body) interaction has to be considered concomitant with the wetting properties of the solvent on the substrate. Another approach is to consider thermodynamic parameters like the Gibbs and Helmholtz free energies, which has been successfully applied to the description of nucleation and growth of atomic thin films and colloidal crystals [26]. However, NP self-assembly is a process far from equilibrium due to solvent flow and local solvent fluctuations [22, 27]. The dynamic problem of solvent evaporation has been addressed with a coarse-grained lattice model to describe the final morphologies comparable to experimental observations [22].

Here we present several examples of studies of self-assembly of iron oxide NPs dispersed in toluene on different substrates. The final morphology can be understood as a result of a two-step process. The first step determines the wetting or dewetting of toluene on the respective substrate. This defines the macroscopic ordering of the NP layers (either continuous layer or islands). The second step is solvent evaporation, where the nanoscopic ordering takes place and the NPs form ordered or disordered structures. The two-step self-assembly process leads to unique morphologies characterized by quantitatively different superlattice coherence lengths and defect structures. Ideally, the two steps are not mutually exclusive events. We investigated these structures in the real space as well as in the reciprocal space via scanning electron microscopy (SEM) and grazing incidence small angle x-ray scattering (GISAXS) at synchrotron sources, respectively.

The first example to be considered is a monolayer thin film of NPs, *spin-coated* onto a Si substrate (sample 'Si'), the scanning electron microscopy (SEM) image of which is shown in Figure 1a. A complete monolayer of NPs is clearly visible, covering the entire substrate surface. The NPs self-assemble into a 2D hexagonal lattice, showing domain boundaries, vacancies and other structural defects analogous to those found in polycrystalline solids. Moreover, in some places it is possible to observe an incomplete second layer forming a system of effectively 1.2 to 1.5 monolayers. The solvent completely wets the surface and the NP assembly takes place during the evaporation. The hexagonal arrangement can arise due to attractive vdW force, which manifests below a certain volume fraction of dissolved NPs with respect to the total volume of the solution.

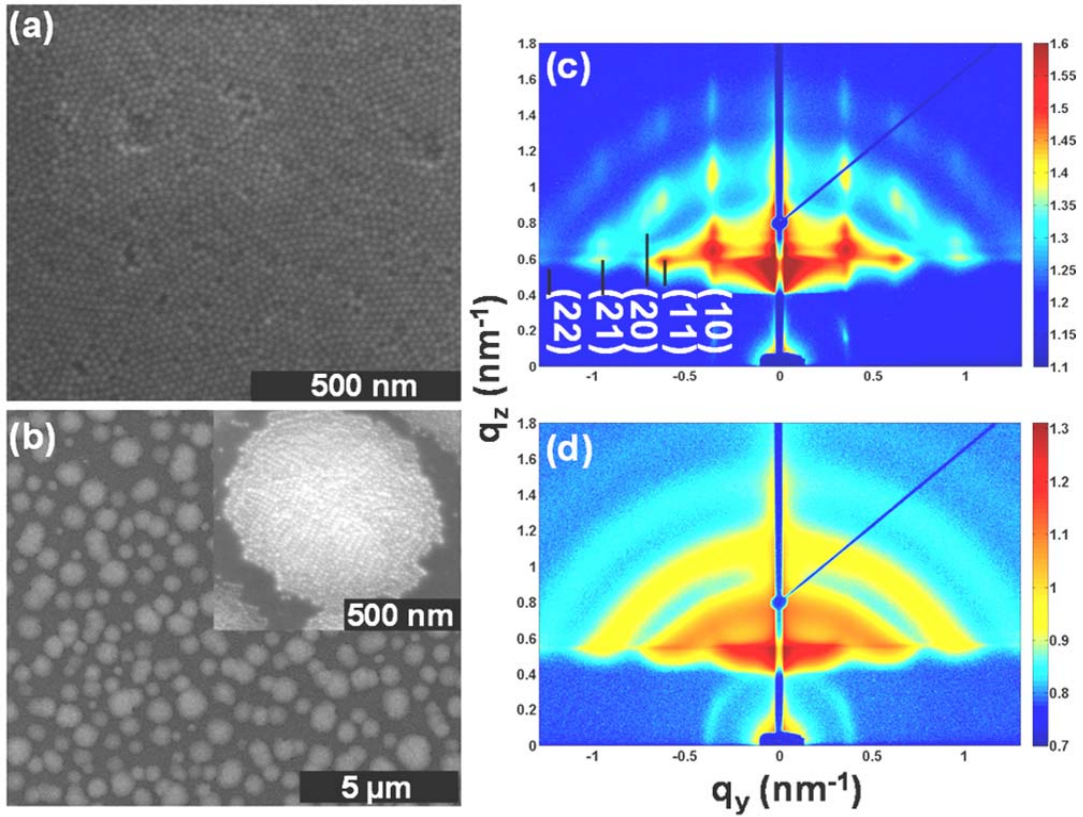

*Figure 1: SEM images of NPs spin-coated on (a) silicon substrate with native oxide (sample 'Si') (b) PMMA coated silicon (sample 'PMMA_ 4P') showing a completely different NP arrangement on the substrates. The corresponding GISAXS patterns are shown in (c) and (d) respectively.*

Although SEM imaging provides us with a useful visual inspection of the sample, it alone is insufficient to address a comprehensive study of long-range correlations of NPs across the substrate. A more powerful characterization technique is needed, providing information on both the in-plane as well as the in-depth structural ordering. These requirements are met by the use of Grazing Incidence Small Angle X-ray Scattering (GISAXS), a surface sensitive x-ray scattering technique, which provides electron density profiles statistically averaged over a large lateral area [28, 35]. By the shallow glancing incident and exit angles of the x-ray beams the surface sensitivity is substantially enhanced [28].

The corresponding GISAXS pattern of NPs on Si, measured at an incident angle ($\alpha_i$) of 0.5º, is shown in Figure 1c, with the intensity plotted on a logarithmic scale coded in color scale shown on the right hand side. Two distinct features can be observed in the scattering pattern. In first place, the intensity is distributed in ring like patterns (although not continuous). At least three such rings can be seen in this pattern. Secondly, on top of the rings high intensity peaks, which modulate the ring intensity, can readily be distinguished. The peaks are extended along $q_y$ and especially $q_z$ directions, where $q_y$ is the lateral component, while $q_z$ is the component of the wave vector transfer **Q** normal to the surface. These appear symmetrically on both sides of the $q_z$ axis along the $\pm q_y$ axis, representing results of the Fourier transform of the in- and out-of-plane electron density variations. The latter basically depend on two factors: the morphology (shape and size) of the NPs and the particle-particle correlations (NP ordering). The ring like pattern is a manifestation of the Fourier transform of the morphology, i.e. the form factor of the spherical NPs.

Simultaneously, the intense Bragg peaks seen at the intersection of rings and streaks (rods) extended in the $q_z$ direction are manifestations of the Fourier transform of the particle-particle correlation function. In particular, the small width of the peaks in the y-direction indicates the presence of a long-range periodic distribution of the NPs over the substrate surface, while their broader line shape in z-direction is due to the small thickness of the NP film. In other words, the high intensity Bragg peaks arise from long-range NP ordering in the GISAXS geometry, where the scattering vector **Q** is of the order of the reciprocal lattice vector (~**G**) of the NP in-plane lattice. The Bragg peaks are an indication of the crystal structure and can be used to determine the NP unit cell structure since in small angle geometry the value $2\pi|\mathbf{G}|^{-1}$ amounts a few tens of nm, which is comparable with the inter-particle distances for the NPs used in this study. The pattern shown in Figure 1c can be assigned to a hexagonal close packed (HCP) lattice with the lattice constant 20.38 nm, which is larger than the NPs average diameter of 18 nm as found from SEM images. Almost the same NP diameter can also be retrieved from the radius of rings in Figure 1c as x-rays are not sensitive to the organic shell and hence measure preferentially the form factor of the iron oxide NP core.

The second system used in this study is illustrated by the image in Figure 1b. It was prepared by spin-coating nanoparticles on top of a Si substrate pre-coated with a few nm of polymethyl methacrylate (PMMA) with 4% solid contents (sample 'PMMA_4P'). In this case the NPs present a completely different ordering compared to the previous substrate. The NPs form islands (mostly disc like) of approximately 1μm in size. The inset shows one of the islands, where the NPs are arranged in a close packed structure. The corresponding GISAXS pattern is shown in Figure 1d. Unlike for NPs spin-coated onto Si, the GISAXS pattern of NPs on PMMA does not show any in-plane Bragg peaks, or rods, indicating that the NPs within the islands are arranged in an amorphous fashion. The ring like structure only arises from the short-range ordering of the NPs (form factor) and does not show any preferred crystallite formation. A feature of particular interest is the agglomeration of densely packed NPs without crystalline structure. The PMMA coated substrate is toluenophobic and forms microscopic droplets with high contact angles. Within the droplets the NPs are strongly bound to the solvent and with evaporation try to agglomerate in order to minimize the surface energy and form a disordered structure. It is clear from these observations that the surface chemical potential indeed influences the self-assembly process.

This will become even more obvious from a direct comparison of two systems presented in the following paragraphs. The first of these systems results from attractively interacting NPs, while the other is dominated by growth formation resulting from the specific form of dewetting behavior. The latter is associated with solvent-substrate interactions, for weakly interacting NPs.

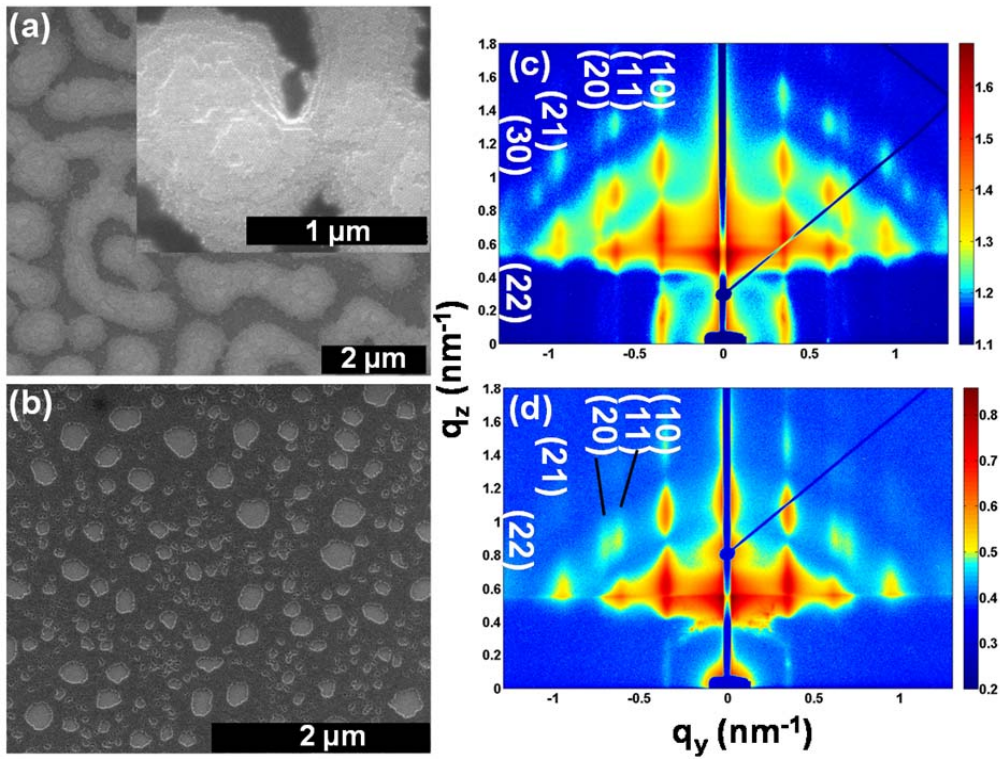

*Figure 2: SEM images of NPs spin-coated on (a) PMMA/MA 33% coated silicon (sample 'PMMA_33P') (b) silicon with 300 nm of silicon oxide substrate (sample 'SiO$_2$'). The inset in Figure (a) shows the spinodal state and the columnar growth of NPs. The corresponding GISAXS patterns are shown in (c) and (d) respectively.*

We consider two systems composed of NPs spin-coated on Si pre-coated with copolymer polymethyl methacrylate/ methacrylic acid (PMMA/MA) with 33 % solid contents (sample 'PMMA_33P') and on Si substrates with 300 nm of thermal SiO$_2$, shown in Figure 2a and 2b, respectively.

The sample on PMMA/MA 33% obviously resulted from a comparably strong interparticle interaction (mostly of vdW type). This can be either explained by a thinner or partial oleic acid shell from fluctuations in fabrication quality or even a full or partial disintegration of the shell due to ageing. Hence, in this specific solution the NPs may have experienced stronger attractive vdW-forces at larger volume fractions of solution compared to the other samples. Attractive interactions between freely moving particles lead to phase separation of particles and solvent by means of spinodal decomposition [22, 27]. Therefore, in this system rather the NP-NP interaction plays the dominant role.

The consequence of the phase separation is observed in Figure 2a, with NPs forming islands being partially interconnected with each other to appear as meander like patterns. The inset in Figure 2a shows a magnified view of one of the islands, composed of several layers of NPs. Remarkably, the NP layers form terraces, which have preferred orientations of crystalline order, although the orientations of the crystallites inside the terraces are different in between any two islands and are probably randomly distributed over the whole substrate. The columnar hexagonal growth seen in the inset of Figure 2a directly indicates the strong NP-NP interaction. The GISAXS pattern shown in Figure 2c is once again representative for a HCP lattice. Note, while the average electron density variation resembles that of a hexagonally arranged monolayer, the formation of randomly oriented and shaped islands add a considerable amount of diffuse scattering around the Bragg peaks, which nevertheless still remain clearly visible and resolvable. For instance, the (30) peak can also be easily identified, confirming the high hexagonal ordering of the NPs inside the islands. Table 1 shows the comparison of intra-planar distance $d_{hk}$ measured from GISAXS patterns and the value calculated assuming a hexagonal lattice of lattice constant 20.38 nm for these NPs. One can recognize a rather good correspondence between the experimental and calculated values from the various peak positions.

*Table 1: Comparison of inter-planar distances measured from GISAXS patterns (2$^{nd}$ column) and calculated assuming a hexagonal lattice (3$^{rd}$ column).*

| Miller indices (hk) | $d_{hk} = \frac{2\pi}{q_y^{hk}}$ (nm) | $d_{hk} = \frac{a}{\sqrt{\frac{4}{3}(h^2 + hk + k^2)}}$ (nm) |
|---|---|---|
| (10) | 17.79 (± 0.023) | 17.65 |
| (11) | 10.3 (± 0.01) | 10.19 |
| (20) | 8.85 (± 0.008) | 8.825 |
| (21) | 6.61 (± 0.004) | 6.67 |
| (30) | 5.87 (± 0.005) | 5.88 |
| (22) | 4.9 (± 0.006) | 5.095 |

As mentioned above, the formation of NP superstructures can be described by the competition of energy terms which account for the complete set of interactions between particle-particle, particle-solvent and particle-substrate [29, 30]. When preparing NP superstructures by spin-coating methods, as in the case of the system presented in this study, further considerations on the wetting ability of the solvent on the substrate should be addressed. As a matter of fact, the colloidal particles are dispersed in a solvent, in this case toluene, and the self-assembly process will also be affected by the solvent evaporation, solvent wetting and its viscosity. The last stage of spin-coating is characterized by evaporation and subsequent thinning of the solvent film [31].

Self-assembly achieved by spin-coating is a dynamic process, where the close packing is mostly determined by the spin speed. By changing the wetting properties of the substrate, it is possible to induce dewetting, occurring simultaneously to evaporation. Dewetting is a process that causes the formation of voids within the uniform solvent film, and to force the surrounding liquid and NPs to move away from them. In this case, the formation of NP superstructures is strongly affected by the solvent evaporation rate [32].

Finally, the image of NP monolayers shown in Figure 2b provides an excellent illustrative example of the effect of dewetting during the formation of NPs superstructures. In this system, the NPs in solution are spin-coated in the same fashion as the system reported in Figure 1a. However, the substrate was changed from Si(100) (with few nm of native oxide layer) to a hydrophobic Si substrate coated with 300 nm of $SiO_2$. The NPs form a monolayer presenting defects and holes, often few hundreds of nm in diameter, as a consequence of dewetting. Still the NPs are packed in a HCP like lattice as confirmed by the GISAXS pattern in Figure 2d. The imperfections in the layer, having a large shape and size distribution, do not affect the Bragg peak positions, however, the absolute intensities of the Bragg peaks are lowered by one order of magnitude with respect to those found in the other GISAXS patterns. This reduction can be accounted for by the penetration of the x-rays into the substrates. At $α_i$ of 0.5º the incident angle is, in fact, above the critical angle of $SiO_2$, so that the x-rays penetrate into the substrate and hence less x-ray photons contribute to the diffuse pattern.

So far we have presented the self-assembly of NPs which is characterized by a very fast evaporation rate of the solvent. The opposite scenario is realized when complete wetting is ensured and solvent evaporation is so slow that it seems virtually non-existent. In this case, the NPs which are free to move within the toluene, given sufficient time, will assemble to an equilibrium position. Such a situation is readily obtained by means of *sedimentation* of NPs, a process by which the solvent is deposited on PMMA coated silicon substrates (either type of PMMA layer) and let to evaporate over an extended time. Complete evaporation takes from few minutes to few hours depending on the volume of toluene, the temperature of the system and the atmospheric pressure conditions.

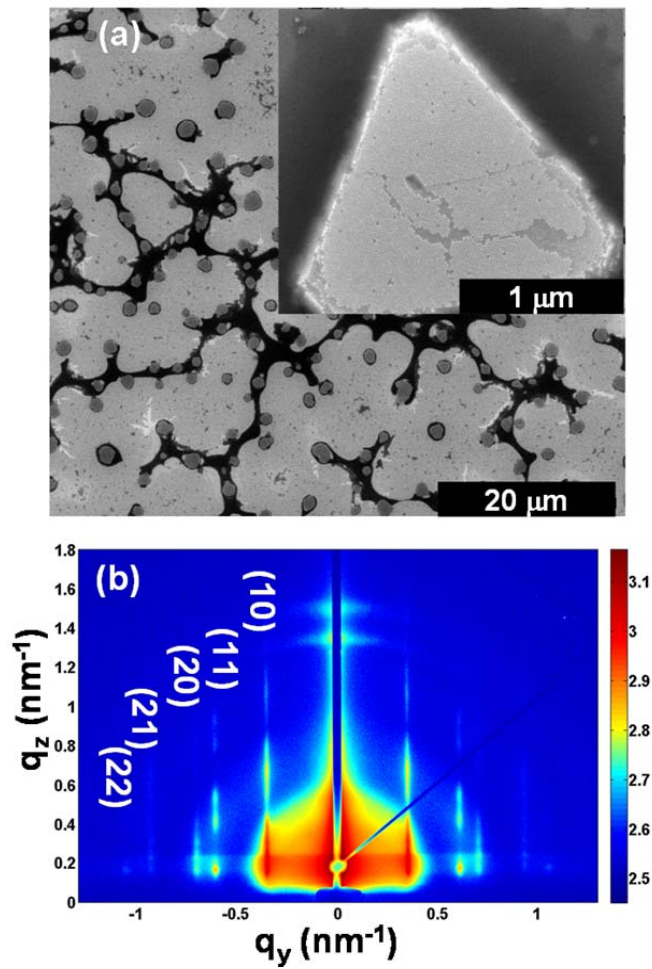

*Figure 3: The SEM images of NPs obtained by sedimentation of NPs on PMMA coated silicon substrates showing (a) islands (NP supercrystals) with random orientations. One observes here the following structures: substrate (light grey), NP supercrystals (medium grey islands) and the residuals from the evaporated solvent (black). The inset shows a single supercrystal with HCP type NP ordering. The corresponding GISAXS pattern is shown in (b).*

Figure 3a and b show the SEM images and the GISAXS pattern of NP self-assembly after sedimentation. The growth should be considered as seed-mediated growth of three-dimensional "supercrystals" or "mesocrystals" [15, 33]. The dark contrast, which appears as bridges between the islands, is due to impurities of the toluene, which concentrate in the remaining solvent before complete evaporation. The image of networks of residual toluene also gives an impression of the complex fluid flow during the evaporation process, driven by the local air-flow, toluene over-pressure and temperature fluctuations over the substrate. As found from the SEM image, the formation of mesocrystals results in this case from another manifestation of a dewetting process during evaporation at very small NP concentration. By examination of the black contrast of the remaining solvent one can easily comprehend the formation of dry cells growing to ever larger ones until coalescing into each other. Hence, the formation of mesocrystals is restricted mostly to areas of dry cells contacting each other, as only then the particle concentration and/or solvent pressure have overcome a critical threshold.

Each supercrystal is single crystalline with facets oriented in differing directions. The nearly triangular shape of the supercrystal shows that the out-of-plane growth direction is primarily along the (0001) axis and further corroborates the assumption that 3D supercrystal growth occurs. From Figure 3a it is clear that the *planar* orientation of the islands is random with respect to each other. Therefore, in the GISAXS pattern the scattered intensity is averaged over all planar orientations. Statistically one can observe all possible lattice planes similar to a powder sample with uniaxial texture such as in graphite, i.e. a 2d- powder pattern.

The GISAXS patterns measured at a glancing angle of 0.1° is shown in Figure 3b. It was not possible to observe any in-plane Bragg peaks at higher angles of incidence. At this very shallow angle of incidence, i.e. below the critical angles for total reflection of x-rays from Si and PMMA, the scattering is dominated by (or originates from) the surface features, viz. the NP assembly, while the x-ray beam does not penetrate the substrate material. Consequently, the shallow incident angle increases considerably the footprint and therefore the contributing scattering volume of the NPs. Hence the best information about the NP ordering inside the islands is obtained at low angles of incidence.

The Bragg peaks are much sharper compared to other GISAXS patterns. Except the (10) peak other peaks have less extension along the $q_z$ axis, indicating a 3-dimensional nature of the islands. As the number of layers perpendicular to the substrate grows, the Bragg peaks become more point-like. Instead of Bragg rods, Bragg spots are seen. The HCP lattice of the NPs is confirmed from the observation of (11), (20), (21) and (30) peaks. The (22) peak is not visible in this case.

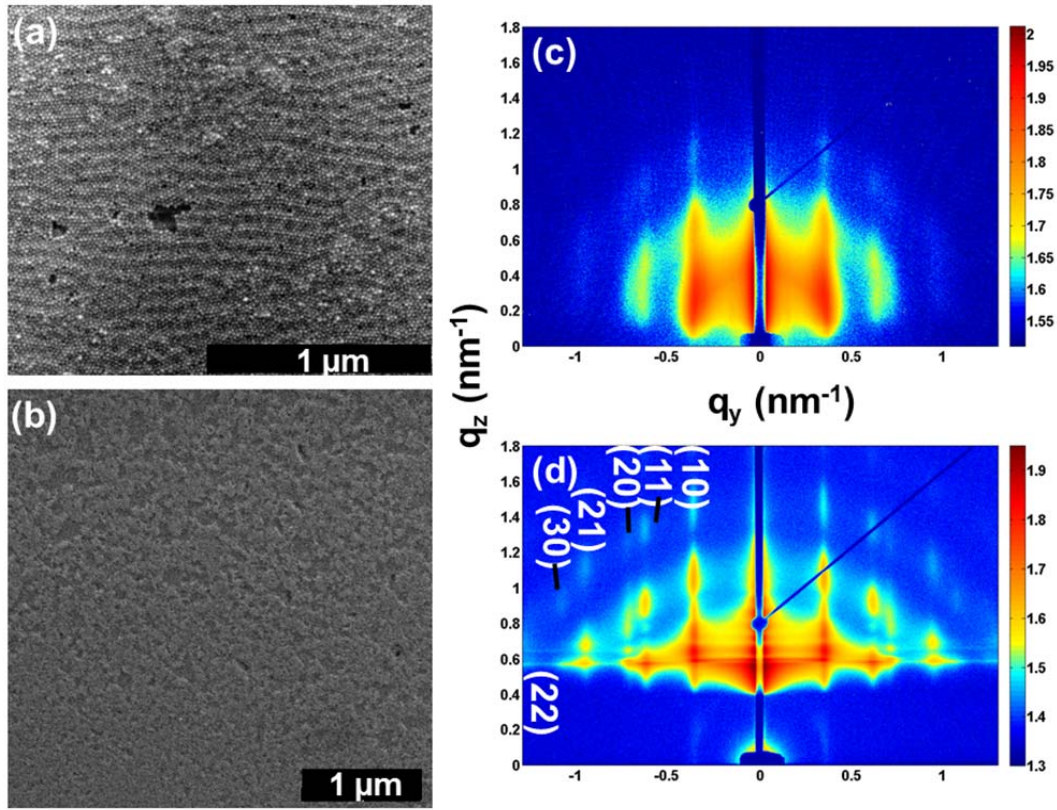

*Figure 4: The SEM images of NPs spin-coated on (a) electropolished aluminum ('Al') (b) a-plane sapphire substrates ('Al$_2$O$_3$'). The corresponding GISAXS patterns are shown in (c) and (d) respectively.*

With regards to the substrates presented up to now, lattice mismatch or the atomic roughness can be neglected as the NP size is much bigger than the atomic size roughness. The surface interaction between substrate and solvent represents the driving force for the growth modes considered earlier, while the NP and substrate interaction plays only a minor role. However, with the appropriate choice of substrate, surface roughness becomes a major player during the ordering mechanism process of NPs. The effect of surface roughness is seen when NPs are *spin-coated* onto electropolished Al and $Al_2O_3$ substrates, as shown in Figure 4a and b. From the SEM image a close packed monolayer of particles can be observed. The Al substrate, electropolished prior to the spin-coating process, has a roughness in the order of a few nm and gives rise to a wavy appearance to the substrate. ($Al_2O_3$ substrate is an insulator and hence one loses resolution at high magnifications in SEM).

Again, by the combination of both GISAXS and SEM imaging one is able to obtain the complete picture of growth modes and self-assembly. The GISAXS pattern in Figure 4c only shows a few broad Bragg peaks. The resolution of this system is complicated by the extreme ductility of the substrate which causes it to slightly bent and deform even during careful manipulation, since the thickness of Al foil is less that 0.5 mm and the discs are very malleable. Nevertheless, the GISAXS pattern of the NPs on $Al_2O_3$ substrate (Figure 4d) clearly evidences a HCP lattice of the monolayer. The (11) and (20) peaks are well resolved. Even at high $q_z$ values it is possible to resolve the (21) and the (30) peaks as well, which was not possible in previous measurements. The narrower peak width along $q_y$ axis indicates larger coherence lengths or larger crystallite sizes. The growth starts with a complete wetting of the solvent plus the NPs. The gradual evaporation of the solvent leads to a hexagonal arrangement of the NPs. The SEM images show that the shape and size of imperfections in the monolayer differ considerably from the ones observed in case of NPs spin-coated on Si.

Having presented a number of case studies of different NP growth conditions, it is clear that the hexagonal close packing is the geometry naturally preferred by NPs for self-assembly, as it either provides the highest packing density in case of dewetting dominated passive deposition of NPs or results from the maximum binding strength in case of attractive vdW-interaction in between NPs. One expects this behavior for systems with negligible entropic effects since these might cause other geometries to be energetically more favorable.

In order to gain further understanding on how the substrate composition and surface roughness affect the degree of packing, we need to look into one-dimensional line cuts at a constant $q_z$ values of the GISAXS patterns presented in this work. The line cuts, shown in Figure 5 for all samples, were taken at $q_z = 0.9$ nm$^{-1}$, with the exception of the NPs prepared by sedimentation (see Figure 5d) taken at $q_z = 0.166$ nm$^{-1}$. The Bragg peaks have been indexed assuming a HCP lattice. For simplicity, the negative $q_y$ axis alone is shown, together with Lorentzian fits to the Bragg peaks. The inverse of the width of the Lorentzian yields the coherence length of the NP crystallites.

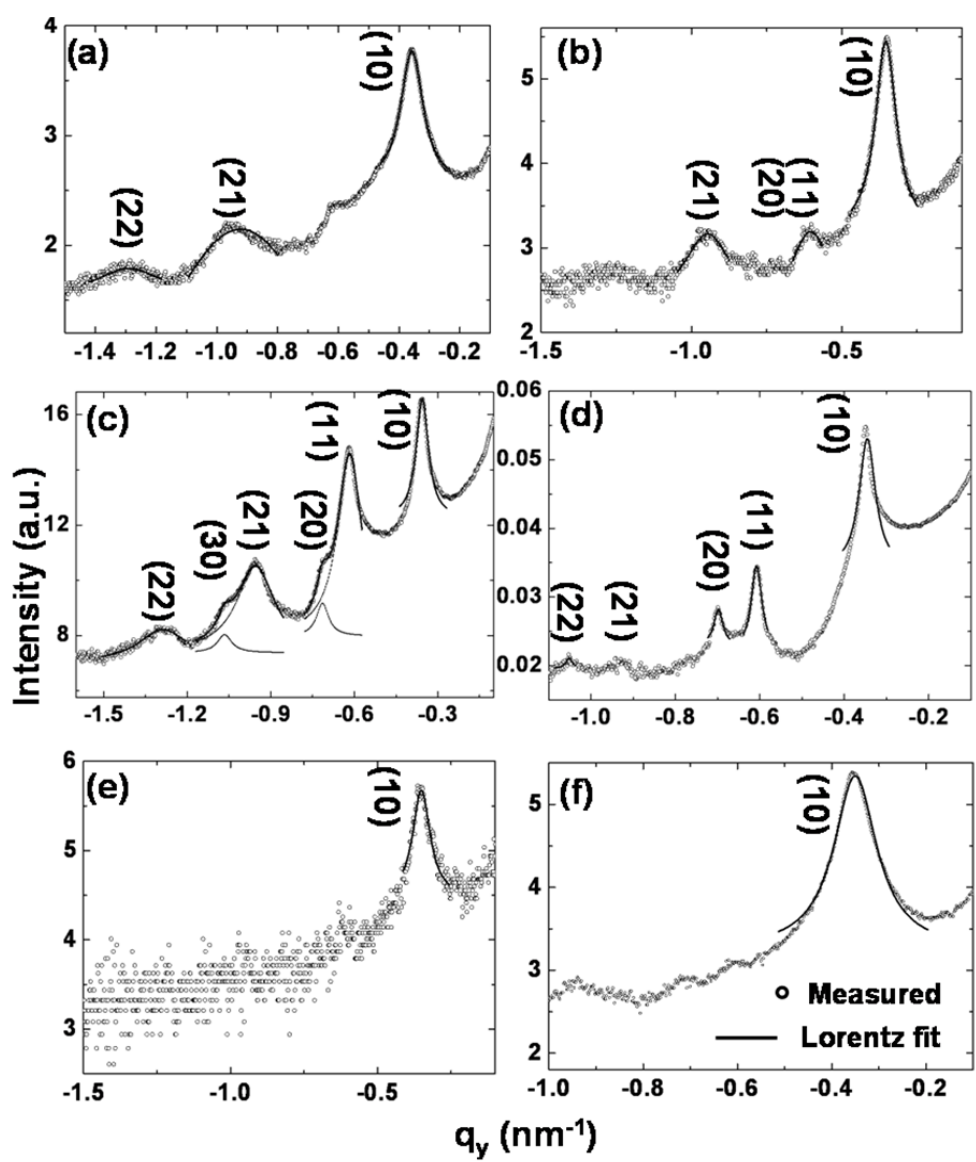

Figure 5: Line cuts at constant $q_z$ for NPs self-assembled on: (a) Si, (b) SiO$_2$, (c) Si+ PMMA/MA (33 %), (d) Sedimentation, (e) Al and (f) Al$_2$O$_3$ substrates. Indexing is given

*according to a 2-dimensional HCP lattice. The open circles are measured data points and the solid lines are Lorentzian profile fits to the peaks.*

The position of the first Bragg peak is similar for all samples. That means that the hexagonal unit cell has the same lattice constant for different growth modes. The main difference between the samples is found in the values of long range ordering of the NPs.

*Table.2. Comparison of intra-planar distance $d_{10}$ (calculated from GISAXS pattern) and the coherence length $\xi_{10}$ of (10) planes (calculated from the Lorentzian fit to the peak of width $\omega_{10}$) for NPs self-assembled on different substrates. The corresponding values for the substrate PMMA (4%) is not included in the table because it has amorphous like structure.*

| Substrate | $d_{10} = \dfrac{2\pi}{q_y^{10}}$ (nm) | $\xi_{10} = \dfrac{2\pi}{\omega_{10}}$ (nm) |
|---|---|---|
| Si | 17.55 (± 0.02) | 70 (± 0.53) |
| Al$_2$O$_3$ | 17.95 (± 0.02) | 90 (± 0.71) |
| Al | 17.9 (± 0.04) | 87.26 (± 5) |
| SiO$_2$ | 17.79 (± 0.01) | 73 (± 0.93) |
| PMMA_4P | 17.95 (± 0.06) | 16.5 (± 1.2) |
| PMMA_33P | 17.65 (± 0.023) | 100 (± 4) |
| Sedimentation | 17.9 (± 0.008) | 282 (± 10) |

Table 2 lists the intra-planar distance $d_{(10)}$ calculated for (10) peak from the Lorentzian fitting and the coherence length of the films from the width of the Lorentzian ($\omega_{10}$) on different substrates. As expected the sedimentation sample has the maximum coherence length of the order of 300 nm. In this case the crystal was formed under an equilibrium condition and develops the maximum ordering.

In all other cases the crystals were formed at a much faster rate and do not have sufficient time to arrange into extended single crystals, rather forming a continuous film with polycrystalline grain boundaries. Curiously, the rougher Al and $Al_2O_3$ substrates, although promoting monolayer growth with higher density of defects, present a slightly higher coherence length than their Si sample counterpart. The coherence length hereby should be considered as an effective parameter, which is the result of the interplay of the various factors during the self-assembly on a particular substrate.

In conclusion, NP ordering differs strongly amongst different substrates, revealing a complex interplay of solvent-, NP, and substrate interaction arising from nanoscale solvent fluctuations due to unique dewetting and evaporation conditions. The morphological resemblance of the self-assembled structures to the growth modes observed in thin atomic films for example Frank van der Merwe and Volmer-Weber growth is quite striking. However, the underlying mechanism is completely different.

The interaction between the solvent and the substrate here is the most dominant of all interactions. The surface interaction defines the wetting ability of the solvent on different substrates. Since in the present case toluene is used as a solvent, the "toluenophobicity" of the substrate determines whether the solution will completely wet the surface or form islands on it. Only in a second step, the ordering or crystallization of NPs takes place. The solvent (toluene) starts evaporating and NPs form a close packed structure. The crystallization process is a slow process. Hence, with spin-coating there is not sufficient time for the formation of large crystals leading to the formation of domains or crystallites. If sufficient time is provided, as in the case of a sedimentation process, it is possible to generate three-dimensional "super-crystals".

The role of the oleic acid shell hereby is to decrease the interparticle interaction, viz. to increase the steric repulsion. Apart from avoiding the NPs to agglomerate in the solution, the thickness of the shell tunes the strength of the NP-NP and NP-substrate interaction during self-assembly and thus modifies the NP film growth mode as described above.

A detailed understanding of NP superlattice growth will provide the ability to control the fabrication of novel nanocomposite materials both as films or as 'supercrystals' with specific physical or chemical properties. Examples are photovoltaic cells, spintronic materials, optical coatings or catalytic systems [12, 37]. To this end the precise structure-property relationship of such NP materials must be established. But to achieve this aim the growth into well-defined structures must be understood. Several open issues remain e.g. the solvent to substrate interaction. Future studies should e.g. quantitatively determine the precise role of "toluenophobicity" of the substrate and very likely also of the NPs.

**Methods:**

**Nanoparticles:** The iron oxide NPs are chemically synthesized and are purchased from Ocean NanoTech LLC Company. The mean diameter is 18 nm with a size distribution of 6.5%. The iron oxide core is surrounded by oleic acid shell with 2 nm thickness. This surfactant prevents the NPs from agglomeration. The NPs are dispersed in toluene solvent and are stored in a sealed bottle. Previous studies on comparable NPs and NPs from this source show that the as-prepared particles consist of two crystallographic phases, viz. maghemite and wüstite [34].

**Samples:** The samples were prepared by spin-coating as described in the literature [34,35]. The concentration of NP in toluene determined the number of layers formed during spin-coating. In order to obtain a monolayer the NP dispersion was diluted by pure toluene solvent in 1:1 ratio. Then 0.01 ml of this diluted solution with NPs was taken from the bottle in an oil and silicon free plastic syringe purchased from NORM-JECT® with EROSA disposable hypodermic needles. Different substrates with typical dimensions of 10×10 mm$^2$ were used. The various substrates used for the investigation are silicon (100) substrate with natural oxide ('Si'), silicon (100) with 300 nm silicon dioxide ('SiO$_2$'), silicon(100) spin-coated with copolymer polymethyl methacrylate/ methacrylic acid (PMMA/MA) with 33 % solid contents ('PMMA_33P'), polymethyl methacrylate (PMMA) with 4% solid contents ('PMMA_4P'), polished Aluminum ('Al') and a-plane sapphire ('Al$_2$O$_3$'). The Al substrates were prepared starting from high purity Al foils (Goodfellow, 99.999%), cleaned by sonication in isopropanol and ethanol for 10 min. The electropolishing of the surface was performed in a 1:3 vol. perchloric acid and ethanol mixture at 10 °C during 10 min, with a constant dc potential of 20 V applied between the sample and a Pt mesh. The Si and SiO$_2$ substrates were purchased from CrysTech. The PMMA_33P and PMMA_4P substrates were prepared by spin-coating PMMA/MA with 33 % solid contents and PMMA with 4% solid contents on Si substrates at 4000 revolution per minute (rpm) for 30 seconds, respectively. After spin-coating it was heat-treated at 80° C in air for 20 minutes on a hotplate (Przitherm PR35 with microprocessor control). The PMMA/MA and PMMA were purchased from Allresist. All spin-coating processes were performed on a commercial spin-coater from SPIN. The NPs were spin-coated in two steps. First, the substrate was spun at 300 rpm for 3 seconds and within 3 seconds the NP solution was drop-cast onto the substrate. This step helps in spreading the solution uniformly throughout the substrate. In the second step, the rotation was increased to 4000 rpm for 30 seconds with an acceleration of 1000 rpm per second. This step determines the thickness of the film. Our effort was to keep the volume of NP solution used and the rotational speed for all the substrates constant. This yields a uniform nominal thickness, irrespective of the arrangement of the particles. After spin-coating, the samples were also heat treated at 80° C in air for 20 minutes for the evaporation of the toluene solvent. The sedimentation sample was prepared in a slightly different way. A PMMA_4P substrate was immersed in a highly diluted solution of NPs. The volume ratio of NP solution to toluene was 1:40000. The beaker containing the substrate and the solution was tightened completely and the solvent was allowed to evaporate slowly overnight through a small opening at the top. All deposition and sedimentation procedures were performed at room temperature.

**Characterization:** A FEI Quanta 200 FEG scanning electron microscope was used for imaging the NPs, which was equipped with an Everhart-Thornley secondary electron detector. The energy was 20 keV with a 30 μm aperture for better resolution. The GISAXS experiments were performed at Hasylab (Hamburg, Germany) beam line BW4 at photon energy of 8.798 keV ($\lambda = 0.138$ nm) [36]. A MAR CCD camera with pixel size 79.1 μm and resolution of 2048 × 2048 pixels was used to capture the two dimensional images. An aperture (0.4 mm × 0.4 mm) and focusing lens system reduced the beam size to 36 μm × 23 μm (horizontal × vertical). The sample to detector distance was found to be 210.44 cm using silver behenate as a calibrant. This gives a q-space resolution of $1.7 \times 10^{-3}$ nm$^{-1}$. The intensities were normalized to the monitor intensity. The geometry of the GISAXS experiment is also shown in Fig 8. The angle of incidence for spin-coated samples was 0.5° and for the sedimentation sample was 0.1°.

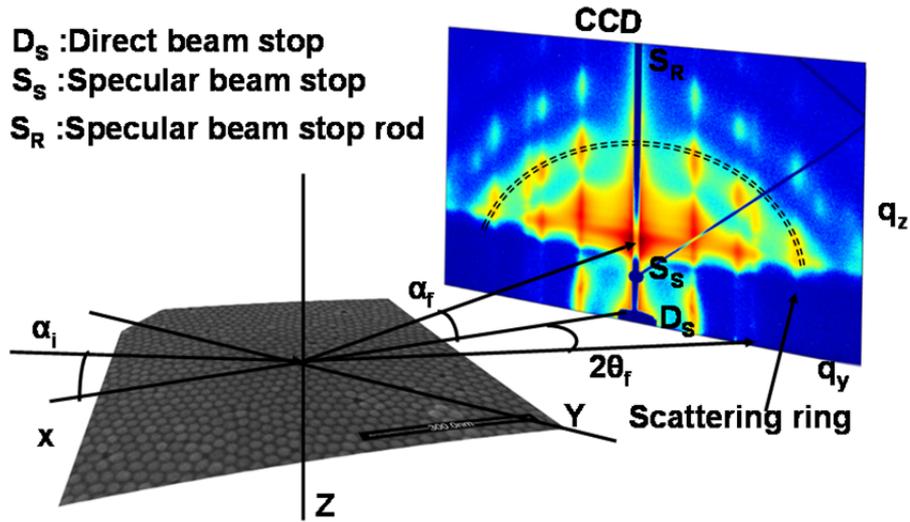

*Figure 6. Geometry of the GISAXS measurement. $\alpha_i$ is the angle of incidence, $\alpha_f$ is the angle of reflection, $2\theta_f$ is the in-plane angle. The specular condition is satisfied for $\alpha_i = \alpha_f$.*

**Indexing the lattice:** The indexing was performed assuming a two dimensional hexagonal lattice. The first Bragg peak is the (10) peak and the lattice constant were calculated from the position of this peak and using the following equation.

$$d_{hk} = \frac{a}{\sqrt{\frac{4}{3}(h^2 + hk + k^2)}} = \frac{2\pi}{q_y^{hk}}$$

Here $(hk)$ corresponds to Miller indices, $d_{hk}$ corresponds to the inter-planar distance, $a$ is the lattice constant and $q_y^{hk}$ is the position of Bragg peak as observed in the experiment.